# Very fast transmissive spectrograph designs for highly multiplexed fiber spectroscopy

Will Saunders[1]
Australian Astronomical Observatory,105 Delhi Road, North Ryde, NSW 2112, Australia

**ABSTRACT**

Very fast (f/1.2 and f/1.35) transmissive spectrograph designs are presented for Hector and MSE. The designs have 61mm x 61mm detectors, 4 or 5 camera lenses of aperture less than 228mm, with just 6 air/glass surfaces, and rely on extreme aspheres for their imaging performance. The throughput is excellent, because of the i-line glasses used, the small number of air/glass surfaces.

## 1. INTRODUCTION

Spectrograph costs have become a strongly limiting factor in many projects for multiplexed fiber spectroscopy, due to increasing fiber numbers and telescope sizes, and resolution and wavelength coverage requirements. They form large fractions of the cost of both the proposed Hector project on the AAT [1], and the Maunakea Spectroscopic Explorer (MSE) project [2]. In both cases, fast camera speeds would allow the use of 4K x 4K detectors, while still giving the required fiber size and resolution. Because the detector and dewar costs are a large part of the total, this could allow a large cost saving. Traditionally, catadioptric designs are used for the fastest cameras, and such a design was proposed for Hector and also 4MOST [3], but catadioptric cameras have many disadvantages[2]. Hence a very fast, efficient, and affordable general purpose transmissive camera design would be nice to find.

Fast transmissive cameras are bound to involve extreme aspheres, or many lenses (or both). Extreme aspheres continue to become easier to make and test, because of the widespread availability of MRF machines and aperture-stitching type interferometric testing. Early versions of this design restricted the extreme aspheres to $CaF_2$ lenses, because these can be diamond-turned to within ~1μm of the required shape and then polished with MRF. $CaF_2$ is also a superb optical material in terms of transparency and dispersion. However, its CTE mismatch with other glasses make it difficult to incorporate into compound lenses, requiring oiled couplings. The availability of i-line glasses, with excellent transparency down to 365nm and matched CTEs, combined with mechanical fine grinding/rough polishing and then MRF, offers an alternative route, and is the one taken in this study.

Spectrograph requirements are somewhat unusual for lens manufactures, in that the requirements for surface accuracy are typically modest (because we do not need to form a very sharp image of the slit), whereas our need for high throughput is paramount. Usually, we are also very sensitive to scattered light, because of the huge range of brightnesses of targets and in the sly spectrum for typical observations.

VPH efficiency profiles allow an efficient design only over a maximum factor 3:2 in wavelength for each arm. This is much less than the typically required wavelength range, which is usually set by the atmosphere and fiber transmission in the UV, and detector technology in the red. Hence modern fiber-fed spectrographs designs are almost invariably multi-armed.

Collimators can used direct-imaging of the fibers, or pupil-imaging [4]. Pupil imaging allows much closer packing of images on the detector, and hence a significant cost saving. However, this comes with significant disadvantages, caused by the image on the detector being an image of the fiber far-field illumination pattern.

[1] will@aao.gov.au

[2] obstructed pupil causing light loss, ghosts and scattering; detector buried in the camera causing operational difficulties, and large beam size causing small grating angles and poor VPH performance at these resolutions.

This has a hole in it (from the telescope top-end obstruction), and depends directly on fiber tilt at input, FRD within the fiber, and the telescope speed and telecentricity, which vary across the field. Hence the PSF is variable, both from fiber to fiber, and from field to field for a given fiber, and some of these variations cannot be modelled. With direct-imaging, the same effects only alter the PSF indirectly, through the speed-dependence of the spectrograph aberrations. Since PSF stability and predictability is paramount for most observations, pupil-imaging was not adopted for the designs presented here.

## 2. HECTOR REQUIREMENTS

The Hector spectrographs will be fixed-format. They will accommodate ~6000 fibers, of as large an aperture as is compatible with the resolution requirements. The resolution requirement is ~1.3Å FWHM[3] over the core range of 370-777nm, with an option of extending this to ~1000nm. This requires a total of ~3300 resolution elements in the core range (allowing 4% dichroic overlap), and so could be accommodated by 2 arms with 4K x 4K detectors with a FWHM ~2.4pix.

Hector will use many weak spectral indices mapped across each target galaxy, and the galaxies are at a range of redshifts. This means that contamination of the spectra by Littrow ghosts [5,6,7] could easily corrupt a significant fraction of the data. Hence it was decided to tilt the gratings, with slanted fringes to preserve the Bragg condition within the DCG, to throw the Littrow ghost just off the detector. This introduces as anamorphism into the design.

AAOmega [7] suffers from complex and variable diffractive blazes, imperfectly understood but clearly caused by the detector and spider in the beam. Hence a decision was taken to have a collimator with unobstructed pupil. An off-axis Schmidt collimator works fine, though it involves a freeform off-axis section of a Schmidt corrector.

The required camera acceptance speed is determined by the fiber size, collimator speed, anamorphism in the collimator and at the grating, resolution and wavelength coverage requirements, and adopted dichroic overlaps. For the values adopted here, this leads to a camera speed of f/1.35.

## 3. DESIGN FOR HECTOR

### Overview

The overall spectrograph layout is shown in Figure 2. The design has an off-axis classic Schmidt collimator including a field lens, with two dichroics to separate the Blue, Red and Infrared arms. The salient parameters for each arm are shown in Table 1.

**Table 1. Parameters for each arm of the Hector spectrograph design**

| | Blue arm | Red arm | Infrared arm |
|---|---|---|---|
| Wavelength Range | 370-588nm | 568-790nm | 763-1050nm |
| Grating angles in air | 13.61° | 19.02° | 19.75° |
| Grating line density | 1237/mm | 1186/mm | 915/mm |
| Maximum camera speed | f/1.348 | f/1.347 | f/1.345 |
| Maximum lens aperture | 200mm | 210mm | 210mm |
| Camera lens mass | 17kg | 14kg | 10kg |

[3] Here and everywhere is this paper, FWHM and resolution refer to a Gaussian fit to the pixellated instrumental PSF, assuming constant noise/pixel, this being an excellent approximation to what astronomers usually measure. For the spectrograph designs presented here, the ratio FWHM is invariably very close to the formal value ($\sqrt{3}D/2$) for the projected fiber diameter $D$ without pixellation or aberrations, and this value is used throughout.

Each arm has a freeform Schmidt corrector on the surface of a prism, which is itself bonded to the VPH grating. There is then an f/1.35 camera, with the field-flattening lens also forming the dewar window.

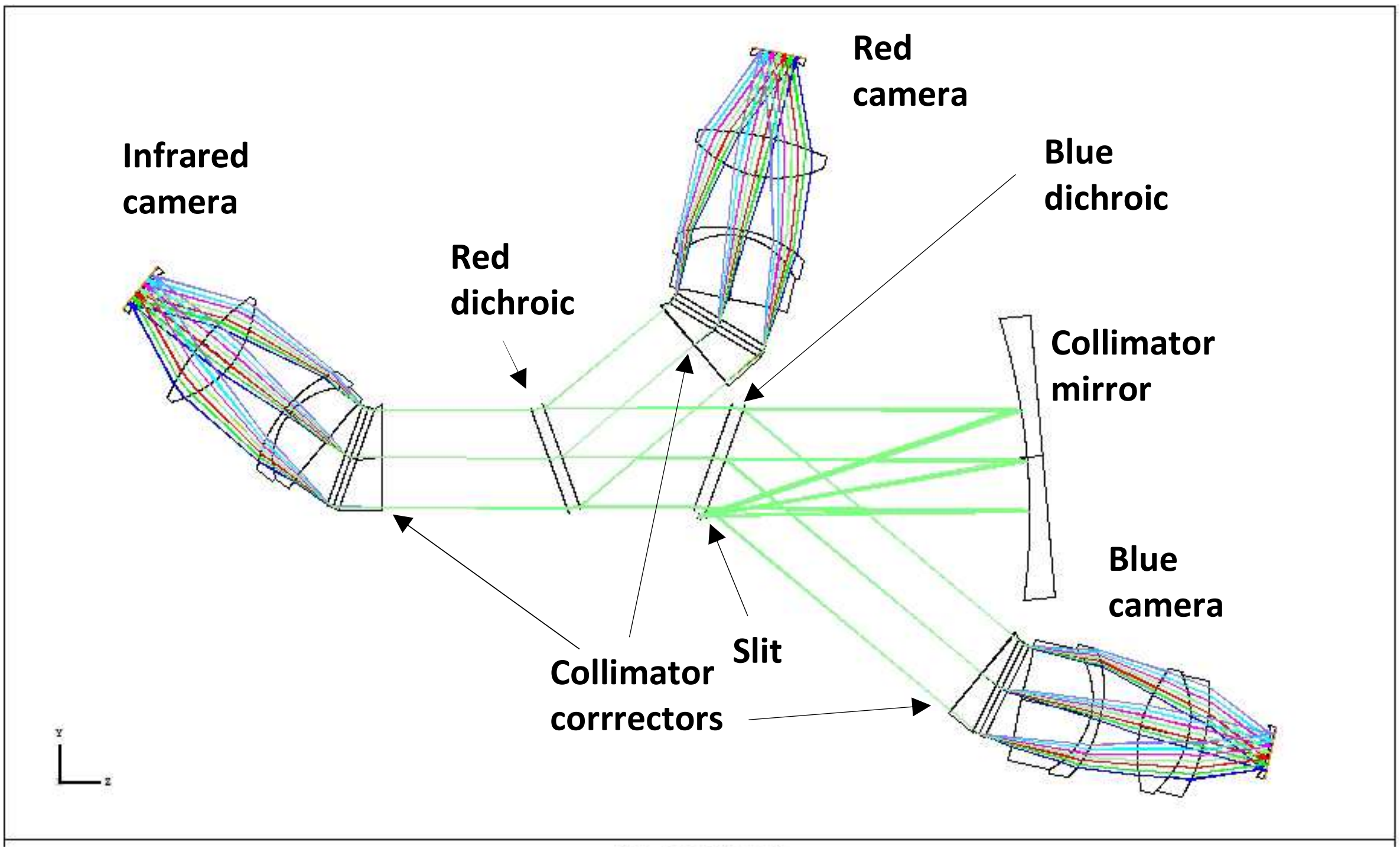

**Figure 1. Overall YZ (horizontal) section for the Hector design. There is a fiber slit, curved both axially and laterally, gelled to a fused silica field lens; a spherical collimator mirror and two dichroics; then each arm has a fused silica VPH grism with the required off-axis Schimdt corrector form on its first surface, and then f/1.35 camera whose last element forms the window for the dewar containing the detector. The collimator speed is f/3.264, and the beam size is 158mm in the larger (spatial) direction, and 155m in the spectral direction.**

### Fibers slit and collimator speed

The proposed 3dF top end [1] has WFNO f/3.427. The design is non-telecentric, but is arranged so that all rays faster than this speed are vignetted at L1, as shown in Figure 2. Hence all fibers are used at a speed no faster than f/3.427. The chosen collimator speed is f/3.264, i.e. 5% overspeed, to allow for focal ratio degradation.

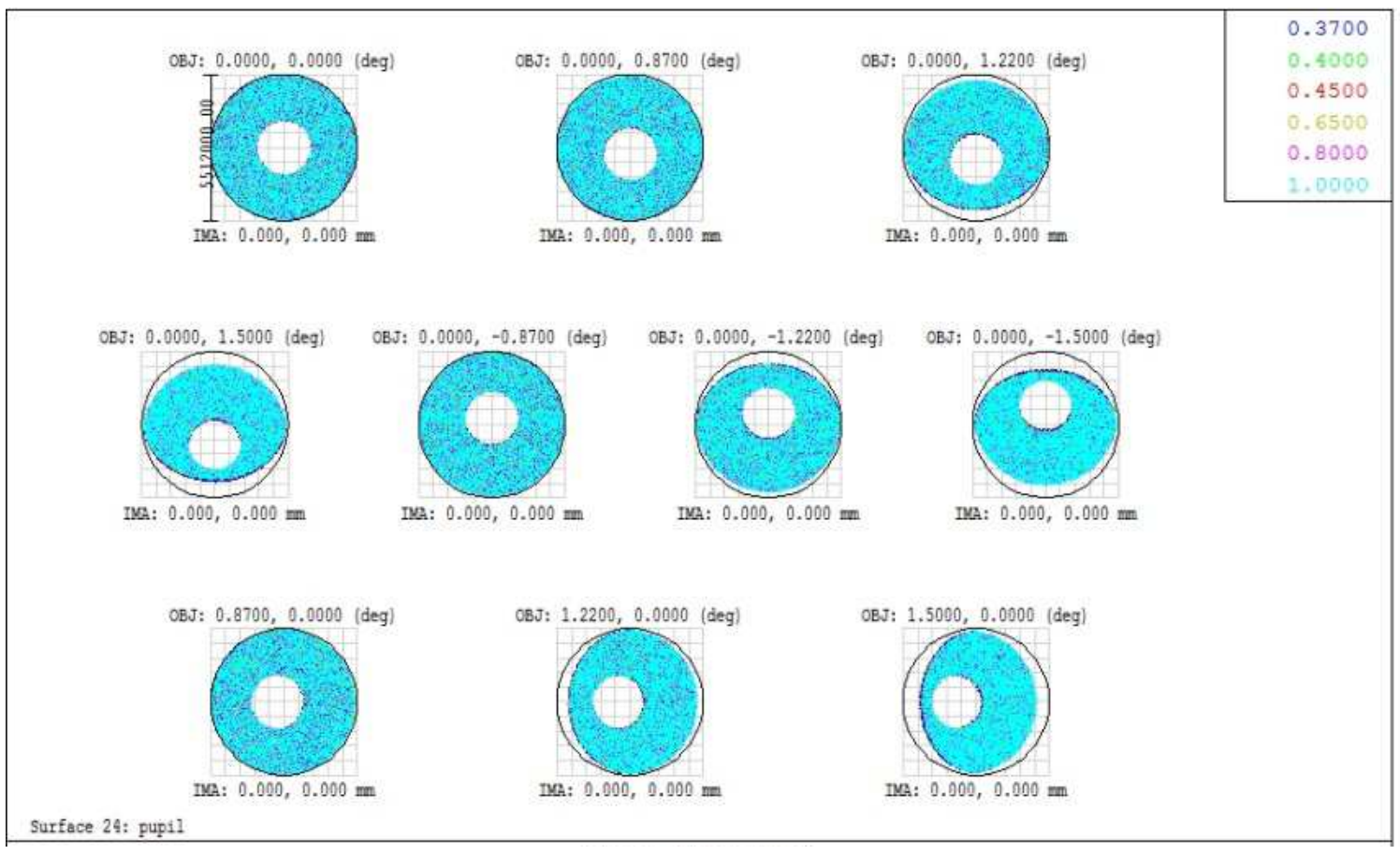

**Figure 2. Fiber illumination across the 3dF field, showing that the fiber acceptance speed required on axis (the circle in the first image) suffices everywhere.**

The nominal fiber size is 100µm, corresponding to 1.54″ on the sky. The fiber slit is 150mm long, and is curved laterally, with a sag of 3.4mm, to minimize spectral curvature on the CCD. It is also curved axially, with a sag of 5.6mm, as part of a Schmidt collimator

**Collimator, dichroics, grisms**

The collimator is an off-axis classic Schmidt. The off-axis design introduces a small defocus between the right-hand and left-hand edges of each fiber, and also a small anamorphism in the beam. The fiber slit It is gelled to a field lens, like AAOmega [7], but of fused silica. The beam size is 158mm in the larger (spatial) direction. There are two dichroics, with AOI minimized. The free-form Schmidt correctors are polished onto the face of fused silica prisms, orthogonal to the beam. Each prism is bonded to a VPH grating, itself on fused silica sub/superstrates. This minimizes the thermal expansion and contraction of the grating, the largest single contributor to thermal shifts of the spectrum on the detector. The gratings bisect the camera/collimator angle, but the prism means they are used away from Littrow configuration, to avoid Littrow ghosts. The VPHs have slanted fringes (by a few degrees) to obey the Bragg condition. The anamorphism in the cameras is ~2%.

**Cameras, dewars, detectors**

The f/1.35 cameras are shown in Figure 3. The Blue camera has two S-FSL5Y/PBL35Y doublets contained in a lens barrel, and a field-flattening lens which also acts as dewar window. The Red and Infrared cameras have a S-FSL5Y/PBM18Y doublet, and a Nikon NIGS4786 singlet in the lens barrel, and a field-flattening lens/dewar window.

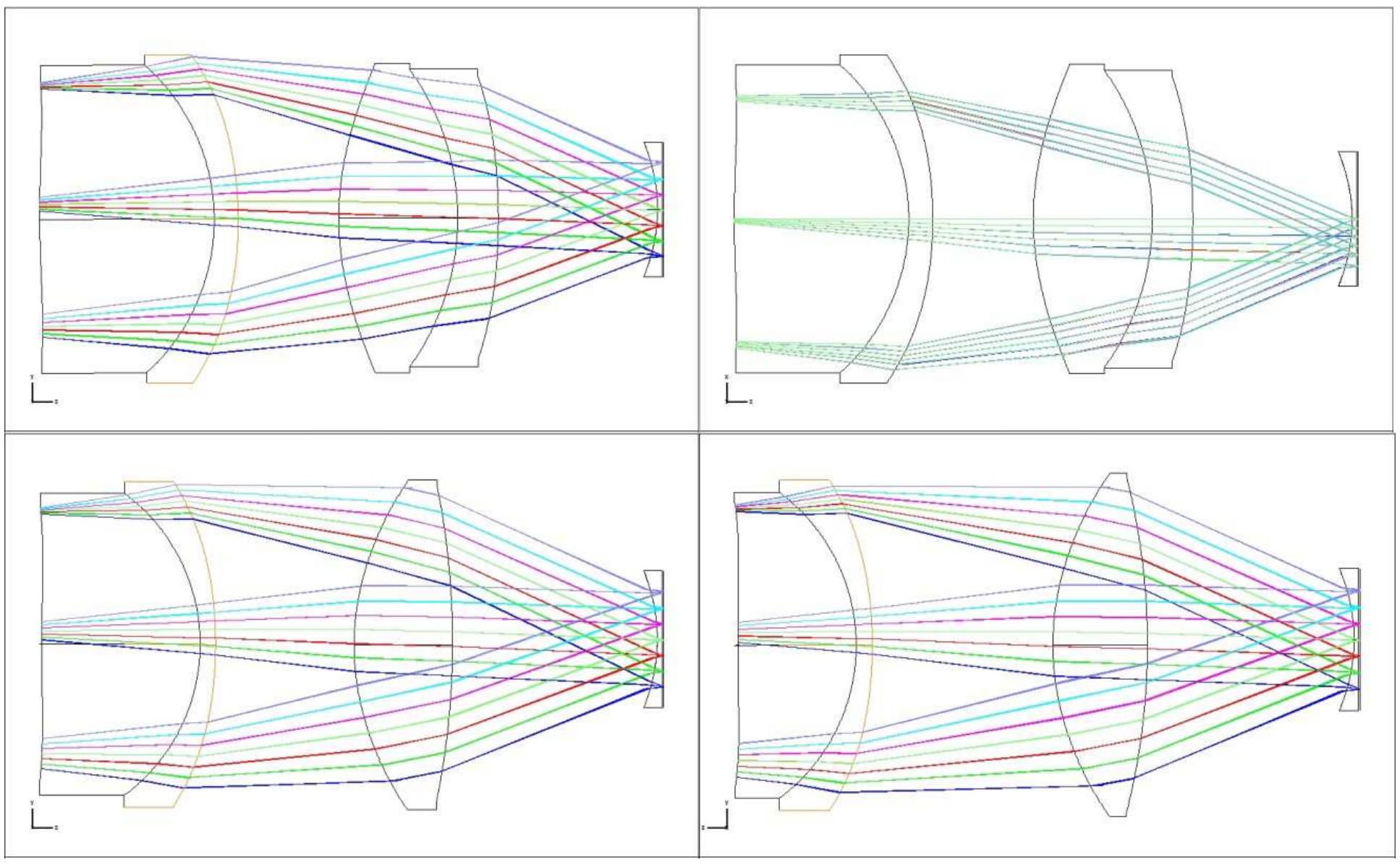

**Figure 3. Top: (a) YZ and (b) XZ sections for Blue camera. The Blue camera consists of two S-FSL5Y/PBL35Y doublets, and a field-flattening lens of AlON. Bottom: YZ sections for (c) Red and (d) Infrared cameras. The Red and Infrared camera consist of a S-FSL5Y/PBM18Y doublet, a Nikon NIGS4786 singlet, and an AlON field-flattening lens. The acceptance speed for both cameras is f/1.35.**

The camera barrels are offset spectrally with respect to the chef ray, by up to 7.5mm. However, the anamorphism means this causes only a few mm increase in the lens diameters. There is also an offset between the two lens groups, arranges such that both groups are slightly rotated in their seats, with maximum decentering from the lens barrel of 300µm. This should be achievable with the AAO's existing optical alignment jig.

The doublets are the key to the imaging performance. As well as being extremely good matches optically, these glasses have almost identical CTEs, allowing them to be glued together. Because they are i-line glasses, they have excellent transparency down to the 365nm. The throughput hit for the doublets in the Blue arm is ~3% at 370nm.

The dewar contains the detector, and has the field flattening lens as a window. The entire dewar is decentered (by up to 2.2mm) and tilted (by up to 0.2°) with respect to the camera barrel. It would be mounted so as to allow motorised adjustment of tip/tilt/piston. Optically, the field-flattening lens can be of any high index and low dispersion glass; the AAO and ANU have used S-YGH51 extensively in the past for field-flatteners, without trouble. However, the Hermes cameras have a significant issue with beta radiation contamination from the particular batch of S-YGH51 used. Tests are continuing at AAO on acceptable substitutes: one promising alternative is AlON (Aluminium OxyNitride), a clear ceramic from Surmet Corp, with adequate transparency, high index and low dispersion, and nearly as strong as sapphire; its principal drawback being that it is difficult to achieve a microroughness better than 5-10nm rms. Sapphire itself cannot be used due to its birefringreance. If no acceptable material can be found, a ~1mm thick fused silica coverslip could be glued to the inner plano surface of the field-flattening lens, to shield the CCD from the radiation.

The dewar is expected to be a modified SpecInst 1100S. These dewars have the advantage that a custom lens can easily be substituted for the standard window; that the CCD/window clearance can be as small as the CCD design allows, and that they can easily be run at the sorts of temperatures optimal for photon-starved spectroscopy (~160K). Shutters will be provided for each camera (either as supplied by SpecInst, or standalone Bonn shutters), so that the exposure times in each arm can be tuned for the competing penalties from cosmic ray hits and read-noise.

The detectors are 4K x 4K format with 15μm pixels, nominally E2V 231-series, tuned for each arm with back-illuminated, deep-depletion and bulk silicon detectors.

Figure 4 shows the full-field spot diagram on the detector, for all fields, wavelengths and configurations. The three arms have different spectral curvature, because they are have different grating angles. However, there is pincushion distortion in the cameras, which allows some slit curvatures without additional detector wastage. A lateral slit curvature of sag 3.4mm puts all the spots on the detector for all three arms.

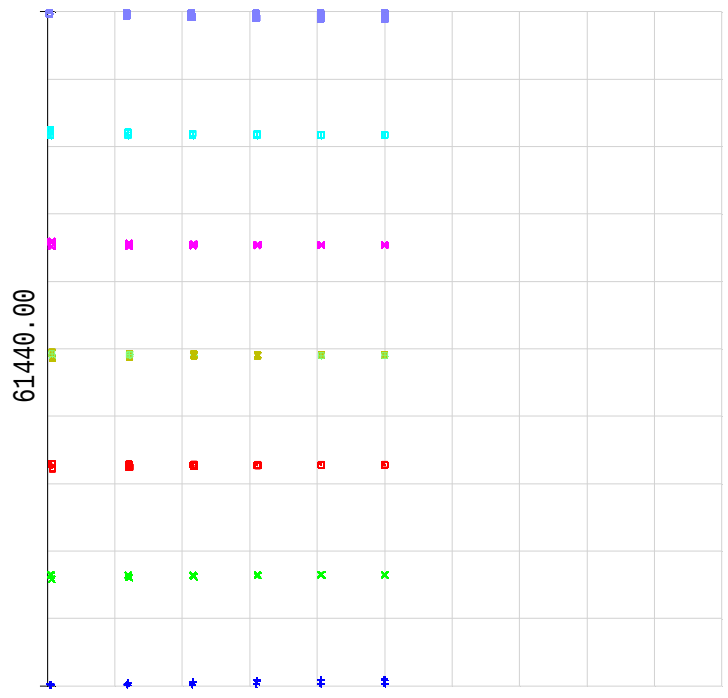

**Figure 4. Full-field spot diagram on the detector, for all fields, wavelengths and configurations. The combination of slit curvature and pincushion distortion allows all spots to be on the detector, despite the different resolutions (and hence spectral curvatures) in each arm.**

## 4. OPTIMIZATION AND IMAGING PERFORMANCE

The optimization was performed with configuration weights {4, 2, 1} given to the Blue/Red/Infrared arm, field position weights {8, 2, ½, 2, 8, 32} given from center to end of slit, and weights iteratively adjusted for each wavelength to achieve approximate uniformity of worst rms image radii. The input beam was apodized according to a simple model the expected level of focal ratio degradation.

Spots diagrams are shown for all arms in Figure 5. The worst rms is ~9μm (in the Blue arm), with average value ~6μm. The Red camera is significantly better than the other two. This easily meets the intended criterion, which is that the rms spot radius should be less than ¼ of the projected fiber diameter.

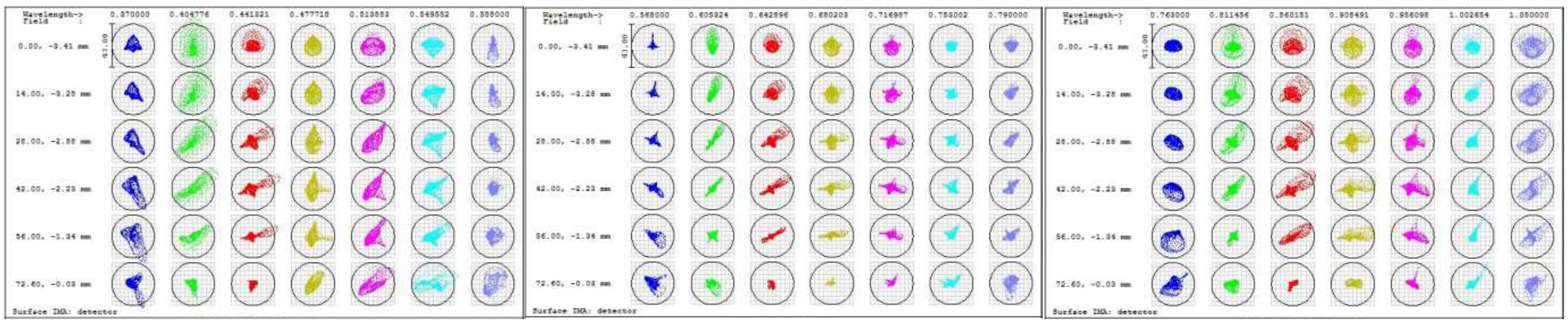

**Figure 5. Spot diagrams for (a) Blue, (b) Red and (c) Infrared arms, for all wavelengths and fields used in the optimization. The circle size is 41μm, equal to the projected spatial fiber size. The rms defocus from the tilt of each fiber is included.**

A preliminary look at tolerances was made using the Zemax tolerance options for decenter (within and between lenses), tilt, and despacing. Surface errors were implemented, for the aspheres only, adding a radial ripple ($8^{th}$ order, Zernike mode 37) to the surface, since ripple is expected to be the dominant surface error. Inhomogeniety was ignored, since the expected contribution is negligible in any case, and all the more so if the aspheres are tested in transmission. All tolerance contributions were given top-hat distributions. The dewar position and orientation were used as a compensators. The maximum amplitudes of each item were iterated until a reasonable image quality was obtained for mildly pessimistic realizations ($+1\sigma$), with comparable contributions from all causes. This gave maximum decenters of ~20μm (both individual lenses and groups), maximum tilts of ~40′, maximum despacings of ~50μm, and maximum rms surface errors of 100nm (so a max PV of ~1/2 wave). The corresponding increase in the average rms radius, given these values, was ~12%, both using rms sensitivity and Monte Carlo methods. This means the expected actual image rms radius is then almost exactly equal to the criterion (worst images ~ ¼ of the projected fiber diameter).

## 5. GRATINGS

Preliminary gratings were simulated in GSOLVER, partly to confirm that the fringe tilt did not cause any significant loss of efficiency (Figure 6). None of the gratings have particularly challenging thicknesses or index modulations. However, there is, as ever, a tradeoff between peak, average and minimum efficiency for each grating, especially for the Blue, which covers the largest fractional wavelength range.

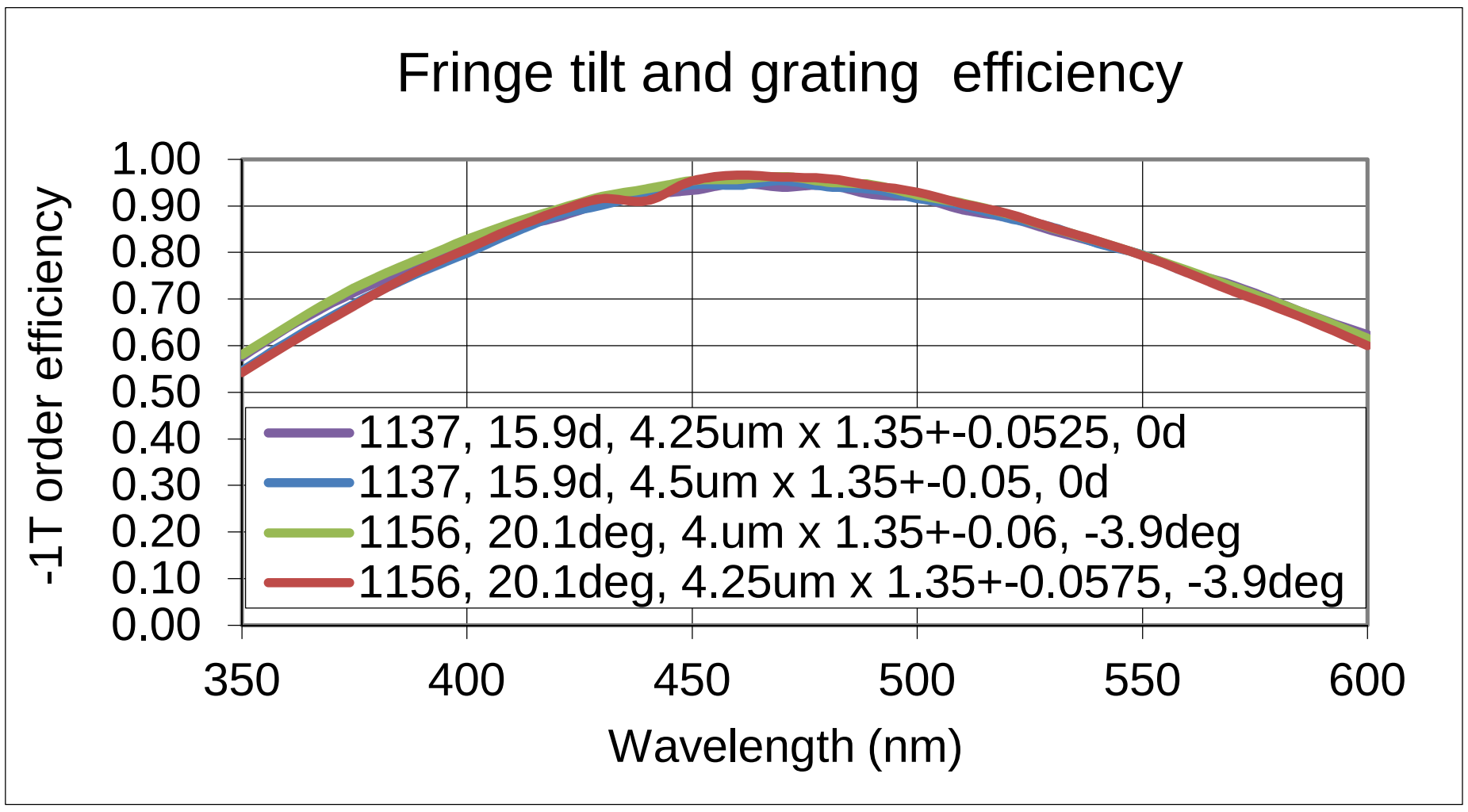

**Figure 6. Four gratings, all with the same dispersion and blaze wavelength. The first two are unslanted, differing in the balance of peak to average efficiency; the last two have slanted fringes, and almost identical performance, though the thicknesses and modulations are significantly different.**

Figure 7 shows the theoretical efficiencies for three gratings modelled for this design. The models are unslanted, but Figure 6 shows that this has negligible effect. The peak theoretical efficiencies are all >90%, and the efficiencies are >70% for all wavelengths 370-1000nm.

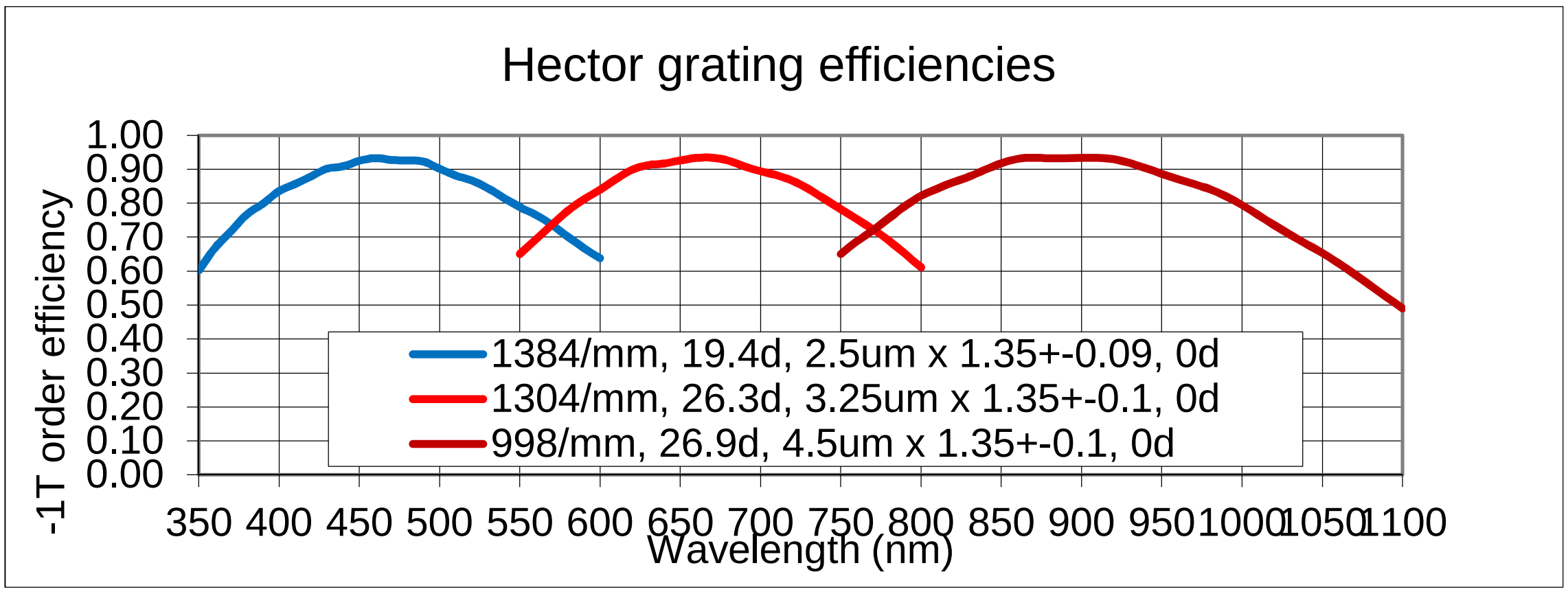

**Figure 7. Nominal grating efficiencies for Hector gratings, modelled without slanted fringes.**

## 6. THROUGHPUT, READ NOISE, CAPACITY

A crude throughput analysis has been carried out, using the gratings as simulated in GSOLVER, with coatings and CCD efficiencies taken from AAOmega where appropriate. The resulting throughput is shown in Figure 8. The peak efficiency in each arm is 70% or better, and the throughput always better than 50% between ~375nm and 950nm.

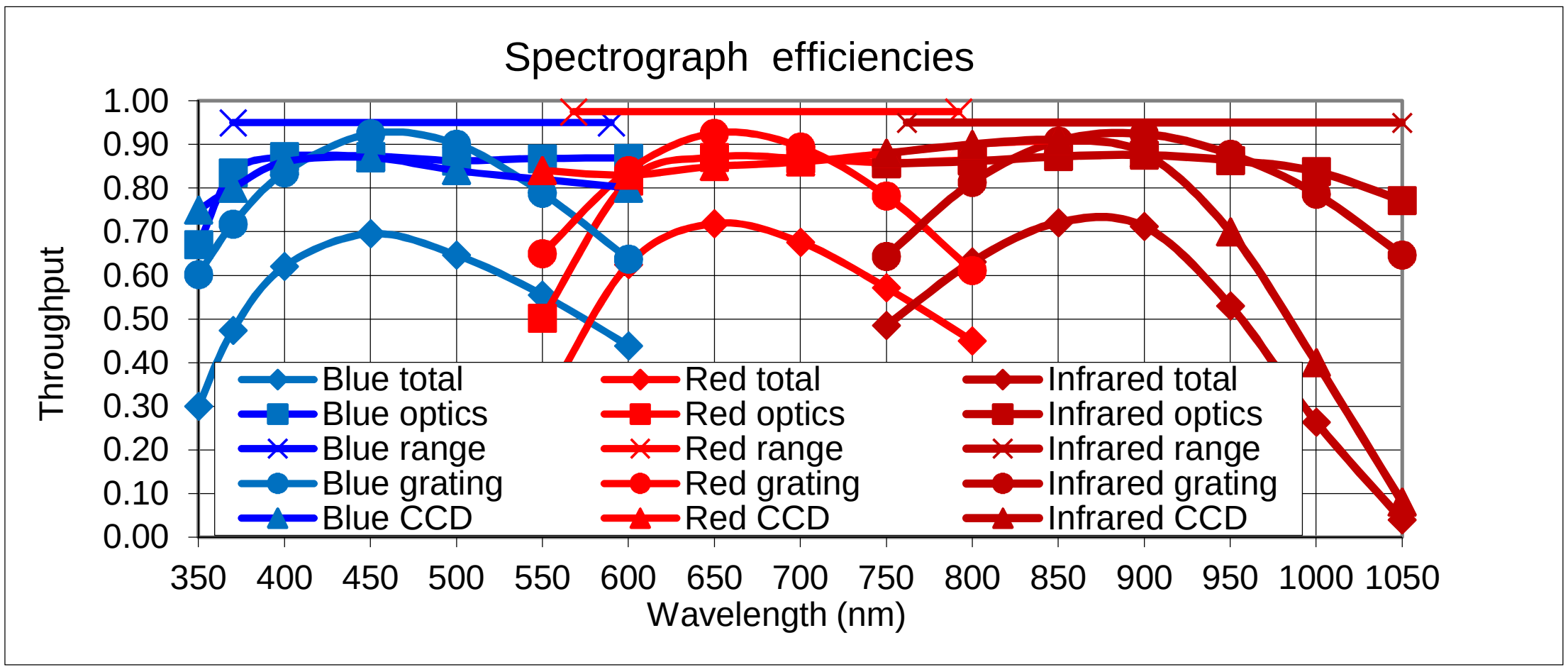

**Figure 8. Preliminary throughput estimate for the spectropgraph. Values for coatings and CCDs have been taken from AAOmega, except the Infrared CCD, taken directly from E2V.**

The throughputs quoted above do not account for read noise, which gives a signal-to-noise penalty that operates effectively as an additional throughput penalty. A remarkably simple and convenient formula for the effective efficiency hit $h$ due to read noise for a close-packed fiber-fed spectrograph is

$$h \approx ½\, r^2\, F_{cam}^2 / (\eta\, s\, \Delta\lambda\, t\, (p/15)^2)$$

Where $r$ is the read noise per pixel in electrons, $F_{cam}$ is the camera speed, $\eta$ is the system efficiency, $s$ is the continuum sky noise in photons/s/µm/m$^2$/arcsec$^2$, $\Delta\lambda$ is the resolution in Angstroms, $t$ is the integration time in hours, $p$ is the (possibly binned) pixel size in microns. The formula is good for any site and telescope, assumes 5% collimator overspeeding, but is only valid to 1$^{st}$ order (i.e. $h$ <<1) For sensible values for Hector ($r$=2.1e$^-$, $\eta$=0.2, $s$=250 photons/s/µm/m$^2$/arcsec$^2$, $\Delta\lambda$=1.3Å, $t$=½ hour, $p$=15µm), this gives effective throughput penalties ~10-13% at all wavelengths.

Cross-talk is not a major issue for Hector, and the spectra would be packed as close as the cladding allows, giving ~1200 fibers per spectrograph.

## 7. A F/1.2 LOW/MEDIUM RESOLUTION SPECTROGRAPH DESIGN FOR MSE

For the proposed 11.25m Maunakea Spectroscopic Explorer (MSE) telescope, the imperative for fast camera speeds is even greater. The nominal spectrograph costs form almost half the total budget. The Low/Medium Resolution (LMR) spectrographs will have a Near-InfraRed (NIR, 1-1.3μm) arm with HgCdTe detectors, limiting the individual detector size to 61mm x 61mm (assuming Teledyne Hawaii 4RG15). Given the cost and procurement issues for these detectors, it is very tempting to aim for a single detector in each NIR camera, and then equally tempting to see if a single 61mm x 61mm detector will suffice for each optical arm. The fiber size will likely be in the range 0.8-1.0″; the required continuous wavelength range is 370-1300nm (with a goal of 360-1300nm), and the required number of spectral resolution elements is 4500-5000. To accommodate this in 3 arms would require f/0.8-f/1 cameras, requiring catadioptric cameras (like MOONS [8]). This would impact efficiency in four ways: the cameras would suffer obstruction losses (the design laid out in [3] providing little benefit because of the very small top-end obstruction); the Blue and Red cameras would each cover a wavelength range 5:3, a very large stretch for the VPH gratings; the gratings would be rather thick (to minimize zeroth order losses, given the small grating angles implied by the large beam required to render the obstruction losses tolerable); and the gratings would need to be tuned for breadth at the expense of peak height. All these problems can be resolved by moving to a design with four arms. A further requirement is that the optical cameras be switchable between Low Resolution (LR, R~3000) and Medium Resolution (MR, R~6000) mode, and a goal that the NIR camera be switchable between *YJ* band (~950-1350nm) and *H*-band (~1450-1800nm). The required camera speed is ~f/1.2, significantly faster than for Hector. However, because the projected fiber diameter is much larger than for Hector (~60μm), the required image quality is also lower, allowing a faster version of the design presented above to be used. This MSE design is much less developed than for Hector, and is presented as a feasibility study to see if the approach is viable.

The MSE telescope design is effectively telecentric, with native speed f/1.926 [9], but this assumes the entire hexagonal segmented primary is used. Since there is almost no light at the fastest speeds (Figure 9), and since the spectrograph costs and difficulties are so great, it makes sense to clip the collimator speed to a more manageable f/2.083, representing a 4-5% light loss, but a 14% reduction in AΩ. High Numerical Aperture (NA) fibers (but still pure silica cored) function acceptably well at these speeds [10]. Remarkably, an f/2.083 collimator is possible with four spectrograph arms, although it means adopting an on-axis Schmidt configuration, with the slit buried inside the first dichroic. Also, the design is extraordinarily compact with little leeway to increase clearances (Figure 10).

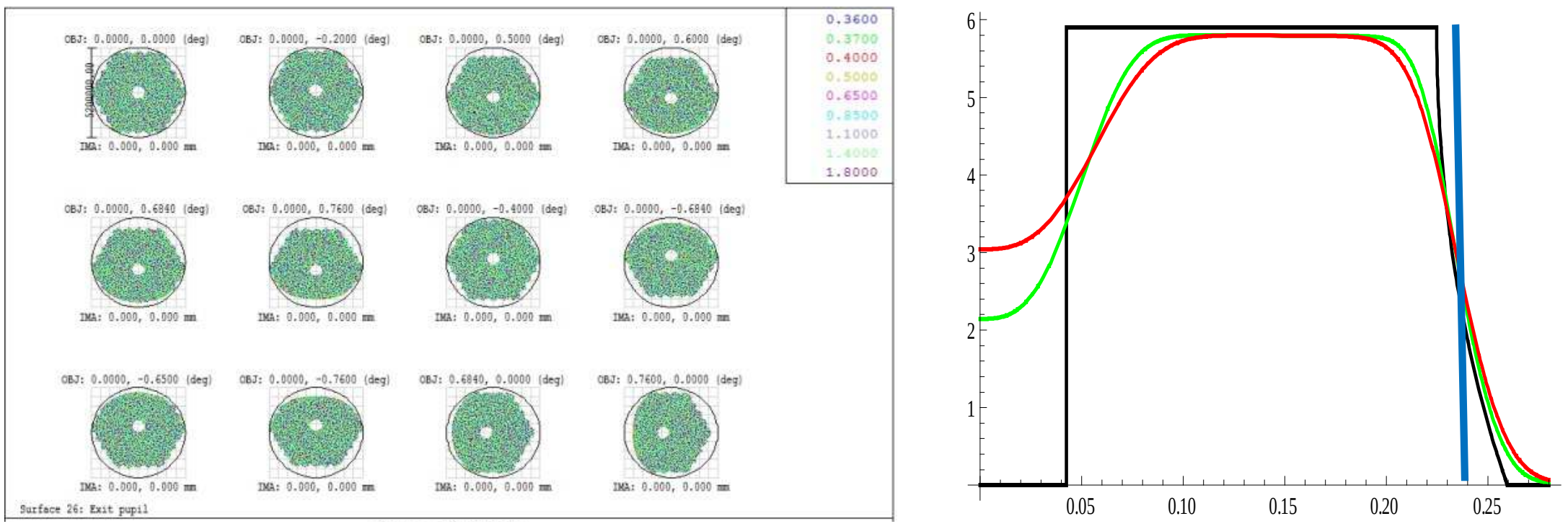

**Figure 9(a). Far-field illumination of the fibers for various field positions. The circle represents the telescope speed on-axis at Zenith (F/1.926); it also suffices at all field positions and zenith distances. Figure 9(b). Radial profile of the MSE beam, without FRD (black), and as modelled with good (green) and bad (red) FRD. The blue vertical line shows a collimator speed of f/2.083 .**

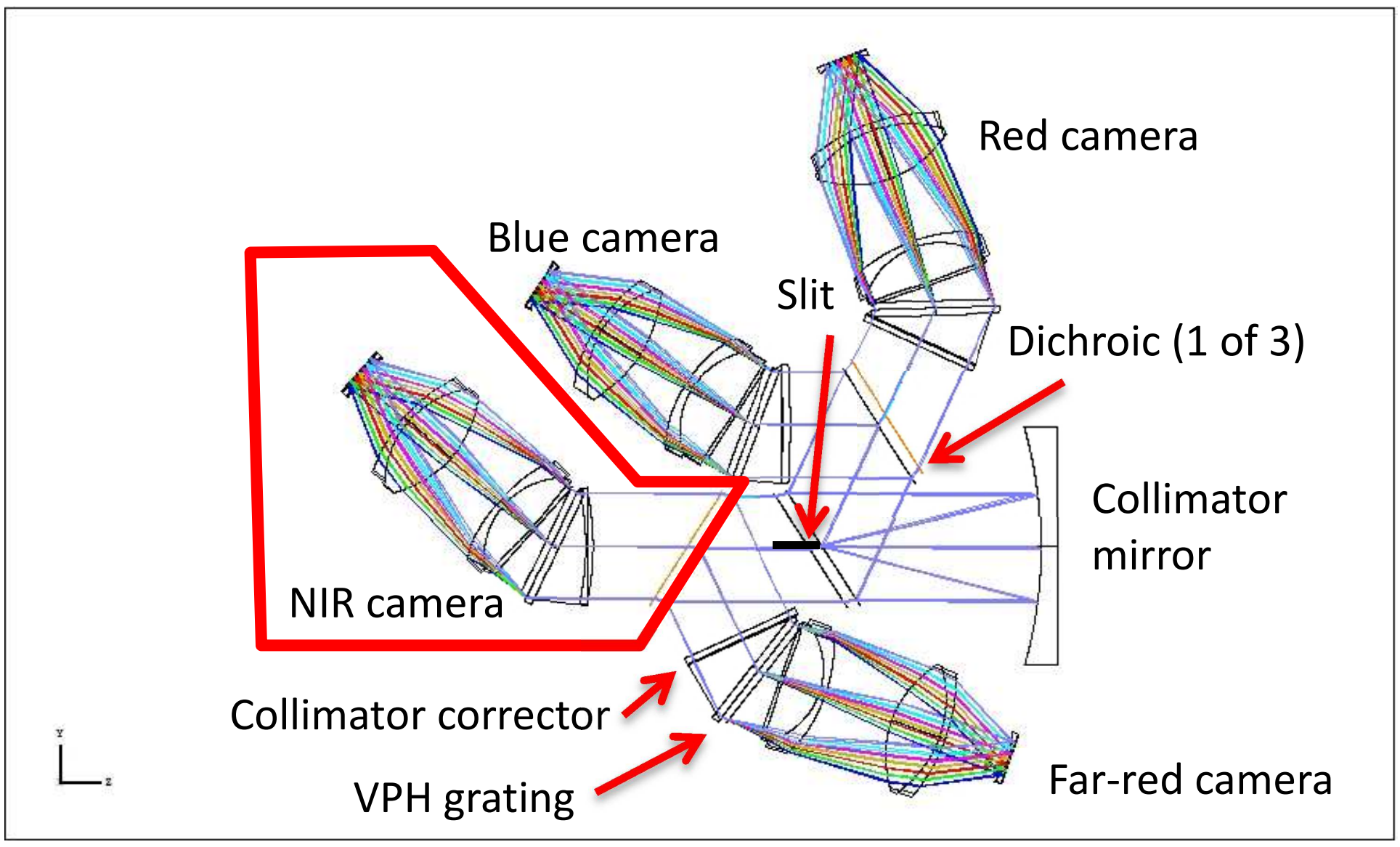

**Figure 10. YZ (horizontal) section through the spectrograph design. The red polygon indicates the nominal envelope of the cryogenic portion of the spectrograph, with the NIR VPH forming the dewar window. It is assumed that the entire spectrograph would be in a deep freezer cabinet with dry $N_2$, to reduce thermal radiation from the slit, collimator mirror, and dichroics.**

The compactness of the design is both a challenge and an opportunity. It raises the possibility of putting the entire spectrograph in a commercial freezer cabinet filled with dry $N_2$, to reduce the thermal radiation from the slit, collimator mirror, and dichroics to acceptable levels; it also makes is straightforward to accommodate the required six spectrographs of the Nasmyth platforms.

The layout of the spectrograph closely follows the Hector design already presented, except that the gratings and collimator correctors are interchangeable between LR and MR modes, on vertical slides. In LR mode, the collimator corrector is a standard Schmidt plate. For MR mode, the Schmidt corrector is formed on the face of a sapphire prism, which is gelled to the fused silica substrate for the VPH grating, with a second sapphire prism gelled to the superstrate. This gives a resolution increase by a factor ~1.8 for MR mode over LR mode. A larger increase could be achieved with a reversed prism in LR mode, but this would increase the required pupil relief, and hence the size and difficulty of the cameras. A larger increase could also be readily achieved by articulating the cameras, but this would represent a significant increase in cost and complexity. Like Hector, the gratings are used away from Littrow, with slanted fringes to compensate, to avoid Littrow ghosts. The salient parameters of each arm in each configuration are shown in Table 2, and the resulting resolutions are shown in Figure 11.

**Table 2. Parameters for the MSE LMR design, for each of the four arms and in each of LR and MR mode**

| | Blue arm | Red arm | Far-red arm | NIR arm |
|---|---|---|---|---|
| LR Wavelength Range | 360-560nm | 540-740nm | 715-985nm | 960-1320nm |
| MR Wavelength Range | 391-510nm | 576-700nm | 738-900nm | 1461-1780nm |
| Grating angles in air | 21.1.8/11.1° | 27.3/17.3° | 27.2/18.2° | 27.5/17.5° |
| LR Grating line density | 1339/mm | 1283/mm | 969/mm | 659/mm |
| MR Grating line density | 2064/mm | 2052/mm | 1599/mm | 737/mm |
| Maximum camera speed | f/1.19 | f/1.18 | f/1.18 | f/1.18 |
| Maximum lens aperture | 221mm | 229mm | 227mm | 223mm |
| Camera lens mass | 14.2kg | 14.5kg | 14.0kg | 13.7kg |

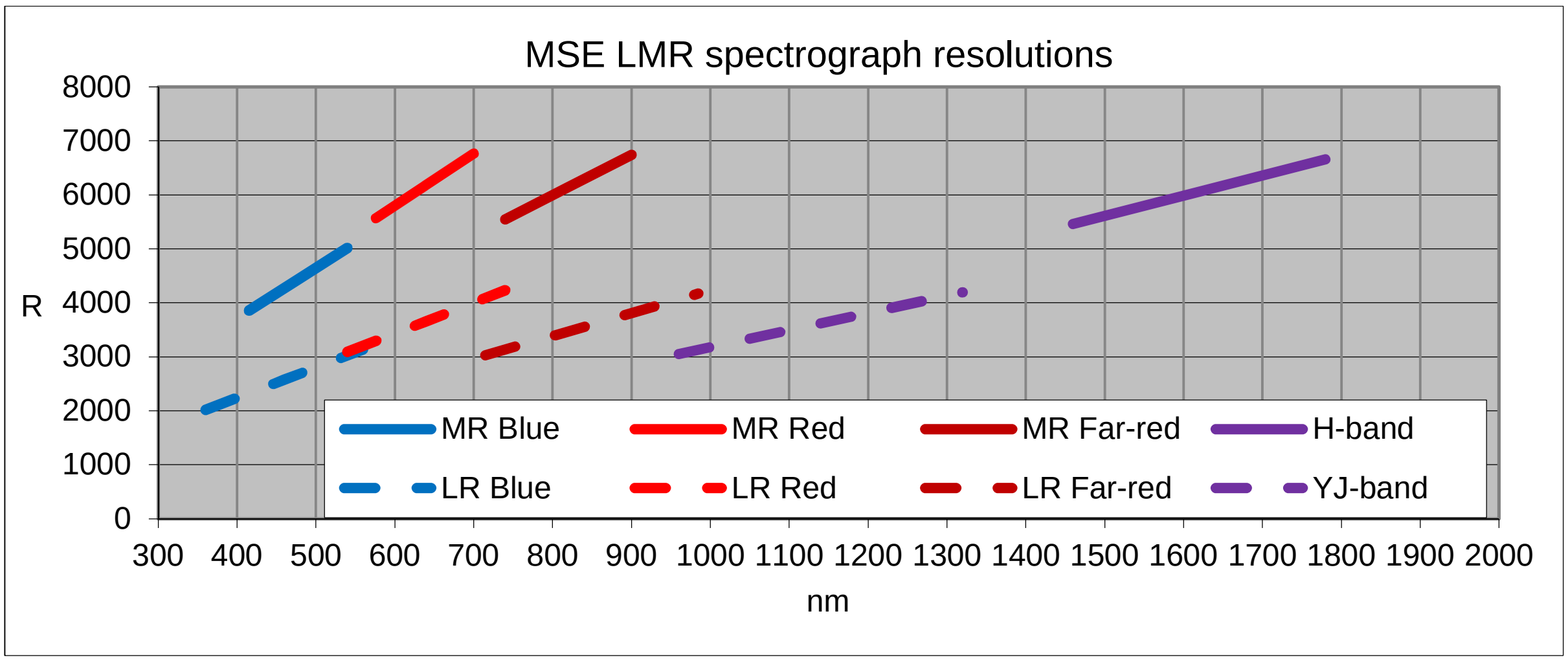

**Figure 11. Nominal resolution for each arm in each mode, assuming 1″ fibers.**

The camera layout is shown in Figure 12. The beam size is 175mm, with the largest lens aperture 228mm. Each camera consists of two S-FSL5Y/PBL35Y doublets and an AlON field-flattener, for which other glasses could be substituted as for Hector.

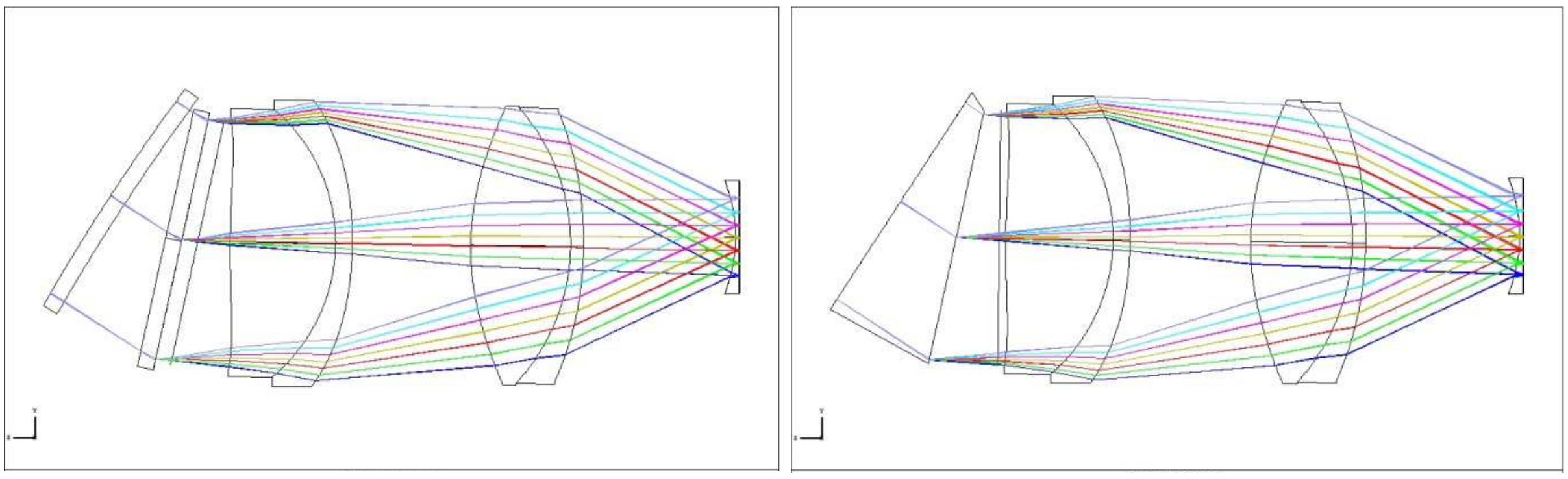

**Figure 12. Blue camera YZ sections for (a) LR and (b) MR mode. To switch modes, the grating and Schmidt collimator corrector is replaced with a grism consisting of sapphire prisms gelled to a fused-silica-mounted VPH, with an even asphere on each external surface. No other motion is required except a <2mm lateral motion of the dewar, and refocus. The camera consists of two S-FSL5Y/PBL35Y doublets and an AlON field-flattener. All lenses are aspheric, with up to 2mm removed material. L1 is gull-winged, and will need testing in transmission. The other cameras are very similar, though L1 is thinner and more aspheric (up to 2.3mm).**

The slit is 101.4mm long, with a lateral sag of 2.4mm, to reduce the spectral curvature on the detector. Unlike Hector, this does not quite allow the differing spectral curvatures to be hidden within the pincushion distortion, but very nearly (Figure 13).

The image quality satisfies the same criterion as for Hector, with the rms radius < ¼ the projected fiber diameter for all fields, wavelengths and configurations (Figure 14). Not surprisingly, there is some loss of quality in the NIR arm, because the camera is optimized for two discrete (*YJ* vs *H*) wavelength ranges.

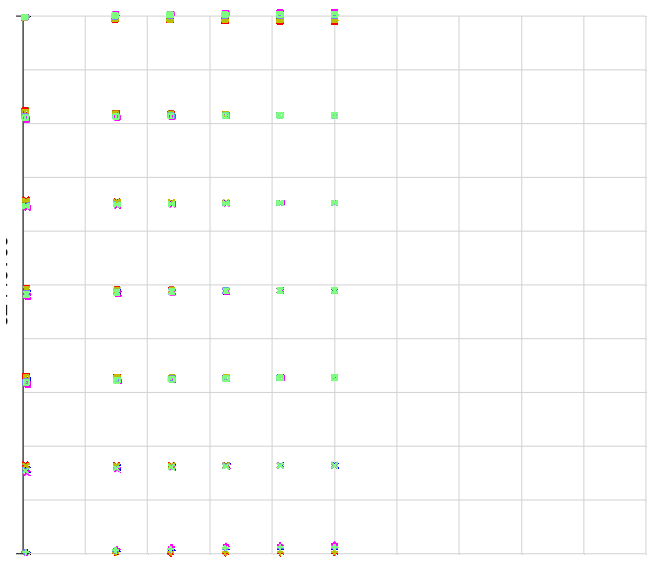

**Figure 13. Image layout on the detector for all fields, wavelengths and configurations.**

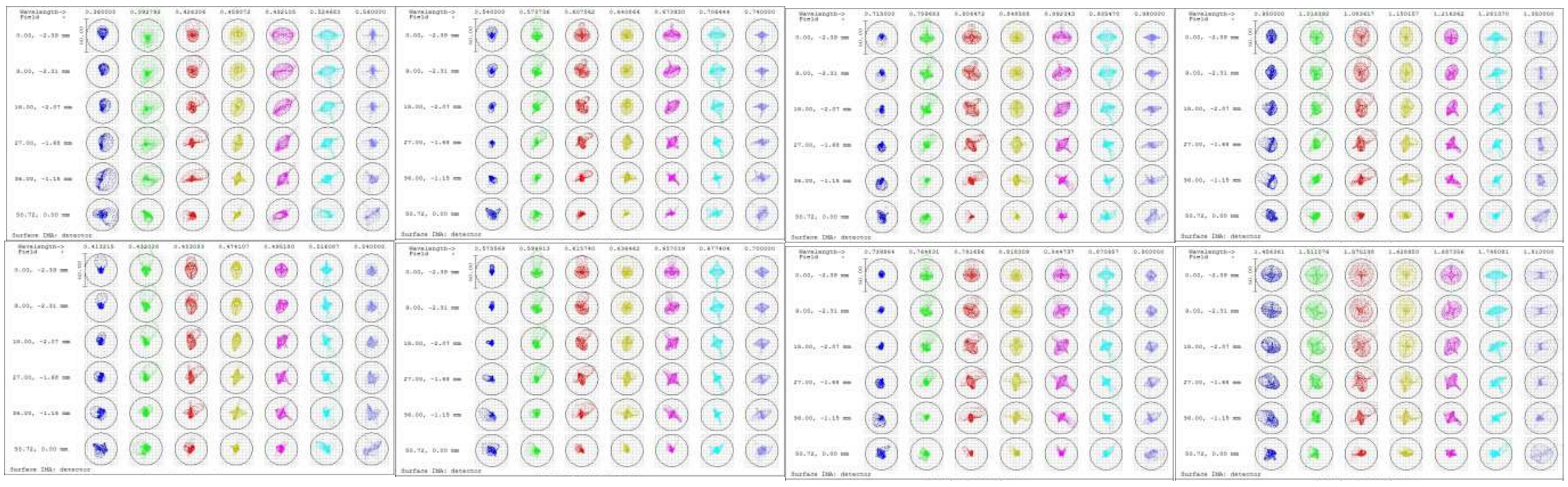

**Figure 14. Spot diagrams for all 4 arms, in (a,b,c,d) Blue/Green/Red/NIR arms for LR mode, and (e,f,g,h) MR mode. Circle size is 60μm, corresponding to the projected fiber size for 1″ fibers.**

The throughput is very good, because having 4 arms means none of them are being stretched too far. Figure 15 shows the nominal throughput for both LR and MR modes. In LR mode, the peak throughput is 64-72% for each arm, and is above 50% everywhere except 360-375nm and 940-965nm. For MR mode, the throughput is above 45% everywhere, with peak 60-68%.

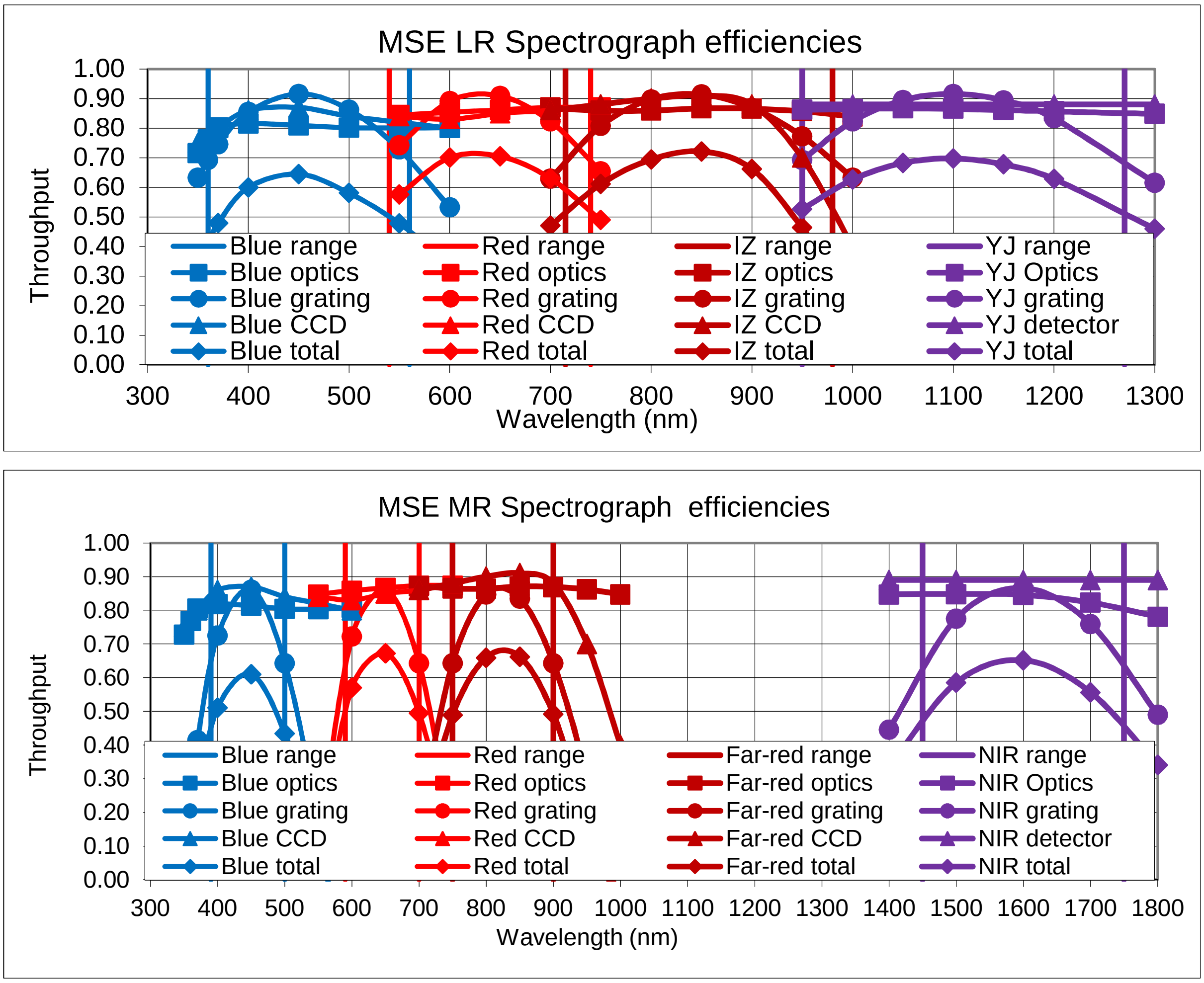

**Figure 15. Nominal MSE spe ctrograph throughputs for (a) LR and (b) MR  mode.**

MSE will have 3200-3500 LMR fibers. With six spectrographs, each with a slit length of 101.4mm and 105µm (1″) fibers, ~550 fibers can be accommodated on each spectrograph, still leaving 3 full pixels between projected fiber images, before accounting for aberrations. Figure 16 shows a simulated image for the end fiber at the blue and red limits (360nm, 1780nm), where the image quality is poorest. The gap between images, including aberrations, is always more than one pixel, ensuring that cross-talk between spectra is very small.

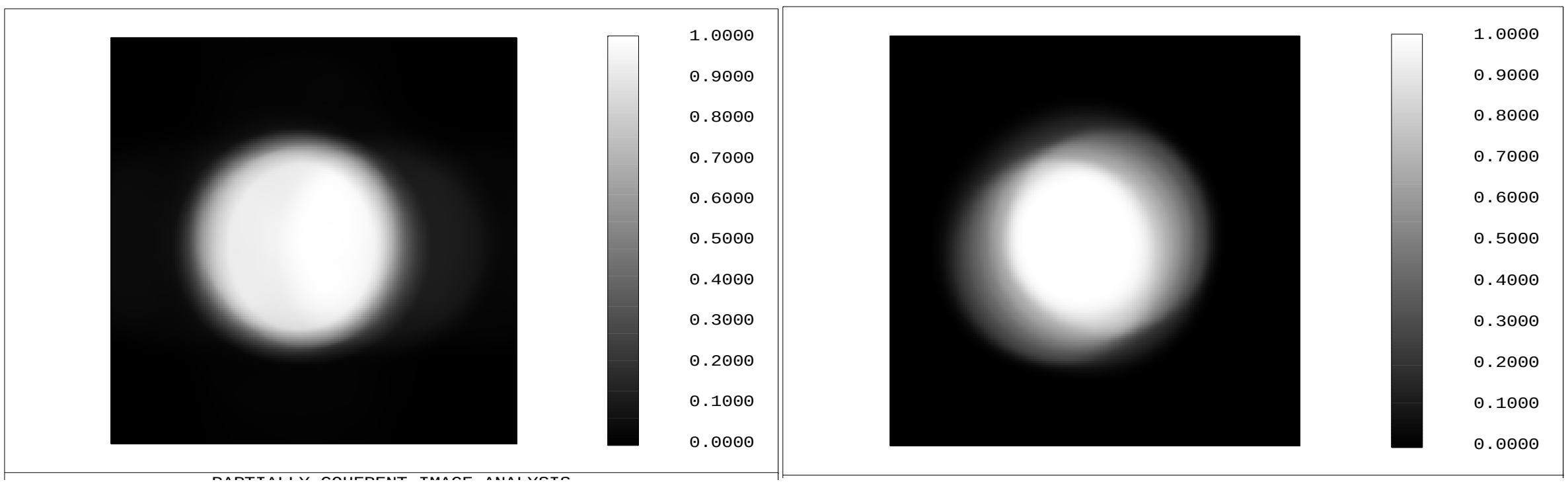

**Figure 16. Simulated images for two of the worst spots, the end fiber at 360nm and 1780nm. The box width corresponds to 210µm at the slit, or 8 pixels at the detector, and the fiber size is 105µm. For 550 fibers/spectrograph, this represents 1.14 pitches. There would always more than one pixel between adjacent images.**

## ACKNOWLEDGEMENTS

Many people have helped with the evolution of this design. Specifically, I thank Chris Hall of QED, Chris Penniman of Nikon, Stephan Klinzing of Asphericon, Jessica de Groote of Optimax; also my AAO colleagues Jon Lawrence, Robert Content, Nick Stazhak, Ross Zhelem, and Julia Bryant, for extensive discussion and input into this design.

## REFERENCES

[1] Bryant, J.J., et al "Hector - a new massively multiplexed IFU instrument for the Anglo-Australian Telescope" Proc. SPIE 9908, 9908052 (2016).
[2] Szeto, K., et al, "Mauna Kea Spectroscopic Explorer design development from feasibility concept to baseline design" Proc. SPIE 9906, 990691 (2016).
[3] Saunders, W. "Efficient and affordable catadioptric spectrograph designs for 4MOST and Hector" Proc. SPIE 9147, 914760 (2014).
[4] Allington-Smith, J., et al. "Integral Field Spectroscopy with the Gemini Multiobject Spectrograph. I. Design, Construction, and Testing" PASP 114, 892-912 (2002).
[5] Wynne, C.G., et al "Ghost images on CCDs" Notes from Observatories 104, 23 (1984).
[6] Burgh,E.B., Bershady,M.A., Westfall,K.B., and Nordsieck, K.H., "Recombination Ghosts in Littrow Configuration: Implications for Spectrographs Using Volume Phase Holographic Gratings" PASP 119, 1069 (2007).
[7] Saunders, W., et al, "AAOmega: a scientific and optical overview" Proc. SPIE, 5492, 389 (2004).
[8] Oliva, E., et al, "The MOONS-VLT Spectrometer: toward the final design" Proc. SPIE, 9908 9908290 (2016).
[9] Saunders, W. and Gillingham, P. "Optical designs for the Maunakea Spectroscopic Explorer Telescope" Proc. SPIE 9906, 9906119 (2016).
[10] Zhang,K-Y., Zhang, J., Saunders,W.,"High numerical aperture multimode fibers for prime focus use",Proc Spie 9912 9912193 (2016).